%% file: main.tex
\documentclass[sigconf,natbib=false]{acmart}

\setcopyright{acmlicensed}
\copyrightyear{2018}
\acmYear{2018}
\acmDOI{XXXXXXX.XXXXXXX}
\acmConference[ExaMPI '25]{Make sure to enter the correct
  conference title from your rights confirmation email}{November 16,
  2025}{St. Louis, MO}
\acmISBN{978-1-4503-XXXX-X/2018/06}

\usepackage[numbers]{natbib}
\usepackage[T1]{fontenc}
\usepackage{graphicx}
\usepackage[algo2e,ruled,vlined]{algorithm2e}
\SetCommentSty{scriptsize}
\SetKwComment{tcc}{\{}{\}}
\SetKwIF{If}{ElseIf}{Else}{if}{}{else if}{else}{endif}
\SetKwFor{While}{while}{}{endw}
\SetKwFor{ForEach}{for each}{}{endfch}

\SetKwRepeat{Do}{do}{while}

\usepackage{hyperref}

\usepackage{amsmath,amssymb,amsfonts}
\usepackage{algorithmic}
\usepackage{graphicx}
\graphicspath{./figures/}
\usepackage{textcomp}
\usepackage{xcolor}
\usepackage{float}
\usepackage{subcaption}
\floatstyle{boxed}

\copyrightyear{2025}
\acmYear{2025}
\setcopyright{cc}
\setcctype{by}
\acmConference[SC Workshops '25]{Workshops of the International Conference for High Performance Computing, Networking, Storage and Analysis}{November 16--21, 2025}{St Louis, MO, USA}
\acmBooktitle{Workshops of the International Conference for High Performance Computing, Networking, Storage and Analysis (SC Workshops '25), November 16--21, 2025, St Louis, MO, USA}
\acmDOI{10.1145/3731599.3767393}
\acmISBN{979-8-4007-1871-7/2025/11}

\begin{document}

\title{Scaling All-to-all Operations Across Emerging Many-Core Supercomputers}

\author{Shannon Kinkead}
\affiliation{%
  \institution{Sandia National Laboratories}
  \institution{University of New Mexico}
  \city{Albuquerque}
  \state{New Mexico}
  \country{USA}
}
\email{skinkea@sandia.gov}
\email{sgkinkead@unm.edu}

\author{Jackson Wesley}
\affiliation{%
  \institution{University of New Mexico}
  \city{Albuquerque}
  \state{New Mexico}
  \country{USA}}
\email{jwesley1@unm.edu}

\author{Whit Schonbein}
\affiliation{%
  \institution{Sandia National Laboratories}
  \city{Albuquerque}
  \state{New Mexico}
  \country{USA}
}
\email{wwschon@sandia.gov}

\author{David DeBonis}
\affiliation{%
 \institution{Los Alamos National Laboratory}
 \city{Los Alamos}
 \state{New Mexico}
 \country{USA}}
\email{ddebonis@lanl.gov}

\author{Matthew G. F. Dosanjh}
\affiliation{%
  \institution{Sandia National Laboratories}
  \city{Albuquerque}
  \state{New Mexico}
  \country{USA}}
\email{mdosanj@sandia.gov}

\author{Amanda Bienz}
\affiliation{%
  \institution{University of New Mexico}
  \city{Albuquerque}
  \state{New Mexico}
  \country{USA}}
\email{bienz@unm.edu}

\renewcommand{\shortauthors}{Kinkead et al.}

\begin{abstract}
Performant all-to-all collective operations in MPI are critical to fast Fourier transforms, transposition, and machine learning applications.  There are many existing implementations for all-to-all exchanges on emerging systems, with the achieved performance dependent on many factors, including message size, process count, architecture, and parallel system partition.  This paper presents novel all-to-all algorithms for emerging many-core systems.  Further, the paper presents a performance analysis against existing algorithms and system MPI, with novel algorithms achieving up to 3x speedup over system MPI at 32 nodes of state-of-the-art Sapphire Rapids systems.  
\end{abstract}

\ccsdesc[500]{Computing methodologies~Parallel computing methodologies}
\ccsdesc[500]{Computing methodologies~Parallel algorithms}
\ccsdesc[300]{Computing methodologies~Massively parallel algorithms}
\ccsdesc[300]{Computing methodologies~Concurrent algorithms}

\keywords{Alltoall, collectives, many-core, MPI, HPC, communication, locality-aware, hierarchical}

\maketitle

\section{Introduction}
Emerging HPC systems continue to scale up through the inclusion of accelerators (e.g., GPUs) and 
increasing core counts in each node. While recent algorithmic focus has favored GPU-aware optimizations, the 
large number of available CPU cores per node presents unique challenges related to inter-process 
communication.  For instance, Sapphire Rapids machines such as Lawrence Livermore National Laboratory's (LLNL) Dane and 
Sandia National Laboratories' (SNL) Amber contain over 100 cores per node split across 2 sockets, and 
further split across 4 NUMA domains within each. Similarly, LLNL's El Capitan leverages AMD MI300A chips, which contain 96 cores per node. These node architectures introduce new 
complexities, amplifying bottlenecks such as injection bandwidth limitations~\cite{gropp2016modeling}.  
Further, the large number of processes at each various level of locality (node, socket, NUMA region, etc) creates disparities in cost between regions of communication.  For example, Sapphire Rapids systems such as Dane and Amber have 14 cores per NUMA region, 56 per socket, and 112 per node, allowing for significant amounts of communication at all levels of the hierarchy, including intra-NUMA, inter-NUMA, inter-socket, and inter-node.

The widely used all-to-all collective creates bottlenecks on emerging systems negatively impacting many different parallel operations. These operations include fast Fourier transforms (FFTs), matrix transpositions, and deep learning frameworks.  The all-to-all operation consists of each process exchanging a unique portion of its local buffer with every other process, requiring a minimum of $\log_{p}$ messages or $s \cdot p$ bytes to be exchanged during an all-to-all operation of $s$ bytes among $p$ processes.  As a result, all-to-all operations have poor scalability, with significant costs associated both with exchange size and process count.  There are a large number of existing all-to-all algorithms, including optimizations for small data sizes that minimize message count as well as those for large exchanges which minimize message volume.  Further, there are a number of existing node-aware optimizations, which aim to reduce message counts and sizes injected into the network, in exchange for increased on-node exchanges. However, the performance of all-to-all operations remains a bottleneck in many applications. 

This paper presents optimizations for all-to-all operations on emerging many-core systems through the following contributions:
\begin{itemize}
    \item Implements existing node-aware all-to-all optimizations, comparing a variety of underlying exchanges within each;
    \item Develops two novel all-to-all optimizations for many-core systems;
    \item Presents performance analysis and scaling studies of both existing and new optimizations across three state-of-the-art Sapphire Rapids supercomputers; and
    \item Achieves up to 3x speedup over system MPI when scaled to 32 nodes.
\end{itemize}

The remainder of this paper is structured as follows. Section~\ref{sec:background} provides background information on the all-to-all operation, details existing approaches, and discusses related works.  Two novel algorithmic approaches are described in Section~\ref{sec:methods}, and Section~\ref{sec:results} presents performance and scalability analysis of both new and existing all-to-all approaches on three Sapphire Rapids systems. Section~\ref{sec:future} discusses the follow on work we are planning. Finally, Section~\ref{sec:conclusions} provides concluding remarks.

\section{Background}~\label{sec:background}
\begin{figure}[h]
\centering
\includegraphics[scale=0.39]{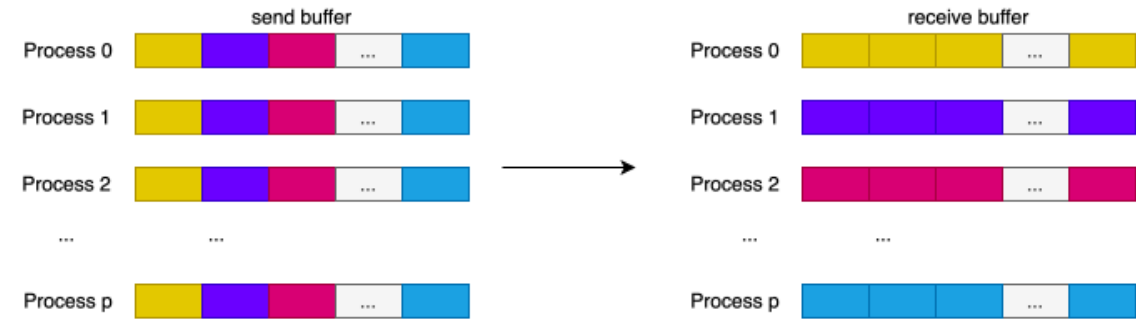}  
\caption{Alltoall between P processes}
\label{fig:all-to-all}
\end{figure}

The all-to-all collective operation, as displayed in Figure~\ref{fig:all-to-all}, exchanges data among all processes within a communicator.  Each process sends an equal-sized portion of data to every other process while concurrently receiving an equal portion originating on each other process, effectively transposing the original data across processes.

There are multiple commonly used all-to-all exchange algorithms.  The Bruck algorithm~\cite{642949} minimizes message counts through a tree-like exchange, requiring $\log_{p}$ steps to exchange data among $p$ processes.  This algorithm requires sending $\frac{s \cdot p}{2}$ bytes during each step, for a given data size $s$.  While this algorithm excels for small data sizes, the data volume reduces its performance for larger values of $s$.

Large all-to-all exchanges minimize the amount of data communicated, with each rank communicating $s$ bytes directly to each of the $p-1$ other processes.  There are multiple approaches for exchanging data, most commonly through either Pairwise Exchange, as detailed in Algorithm~\ref{alg:pairwise_exchange}, or the non-blocking approach detailed in Algorithm~\ref{alg:nonblocking}.

Pairwise Exchange consists of $p-1$ disjoint steps of communication.  At step $i$, process $p$ sends corresponding data to process $p+i$ and receives data from $p-i$, typically through blocking APIs such as \texttt{MPI\_Sendrecv}.  This approach limits network contention and queue search overheads with only a single exchange at any time.  However, large synchronization overheads can occur with pairwise exchange, as at any step $i$, if process $p-i$ is not yet available, process $p$ must wait idly.  

\input{algs/pairwise}

The non-blocking exchange consists of initializing all sends and receives with non-blocking communication, minimizing synchronization costs.
However, this approach can yield significant overheads associated with queue search and network contention at large scales.  

\input{algs/nonblocking}

All-to-all exchanges have been further optimized for recent generations of supercomputers, which have symmetric multiprocessing (SMP) nodes.  Hierarchical collectives gather data among a leader per node before performing an all-to-all exchange among all leaders~\cite{traff2014mpi, zhu2009hierarchical}.  Multi-leader collectives extend the hierarchical approach with multiple leaders per node.  Node-aware aggregation alternatively aggregates all data to be communicated with another node, and then redistributes data within the node~\cite{traff2019decomposing} (Note: A paper by the same name was published in CLUSTER, but only the extended ArXIV version contains the all-to-all algorithm).  All of these existing optimizations are detailed further in Section~\ref{sec:methods}.

\subsection{Related Work}
Additional all-to-all operations have been explored in a variety of contexts.  A batched all-to-all~\cite{namugwanya2023collective} combines pairwise exchange and non-blocking approaches, communicating a batch of messages at each step, in an effort to balance the synchronization costs of pairwise exchange with the contention and queue search overheads of non-blocking.  All-to-all exchanges have been further optimized for GPU-Aware communication, through optimizations to IPC communication~\cite{chen2022highly}, hierarchical optimizations~\cite{zhou2022accelerating}, and compression~\cite{zhou2022accelerating}.  Node-aware aggregation techniques have been widely explored within all-to-all operations and well as its variable-sized counterpart, MPI\_Alltoallv~\cite{khorassani2021adaptive, kefan}.  Intra-node shared memory optimizations have also been explored throughout collective operations~\cite{intra_node_collectives}.

Locality-aware optimizations have been explored in many other contexts, including a variety of collectives~\cite{locality_aware_allgather,locality_aware_allreduce}, collective IO~\cite{node_aware_IO}, and irregular communication~\cite{bienz_node_aware_spmv,collom_neighbor_loc,ygm}.  Topology-aware collectives have further been explored, extending optimizations to the network topology~\cite{topo_aware_alltoall,topo_aware_dkpanda,topo_aware_gropp,topo_aware_hierknem,topo_aware_ma,topo_aware_reordering}.

This paper aims to build on the existing body of work on all-to-all collective optimizations through multiple contributions. First of all, we compare the existing node-aware, hierarchical, and multileader all-to-all implementations, including analyzing the underlying data exchanges. We extend the Locality-Aware collectives presented in \cite{locality_aware_allgather,locality_aware_allreduce} to the all-to-all collective operation, as well as introduce the Multileader with Node-Aware all-to-all algorithm, and analyze the performance and scalability of these new operations in comparison to the existing all-to-all optimizations.

\section{Methods}~\label{sec:methods}

All-to-all exchanges can be optimized on emerging architectures through node- and locality-awareness, differentiating between local and non-local exchanges.  Each of these approaches, described in Algorithms~\ref{alg:hierarchical} to~\ref{alg:hierarchical_locality}, requires performing an all-to-all exchange on an MPI sub-communicator.  Each of these all-to-all operations could use any underlying implementation, such as Bruck, pairwise exchange, or non-blocking.

\subsection{Hierarchical}
\begin{figure}[ht!]
\centering
\includegraphics[width=0.5\textwidth]{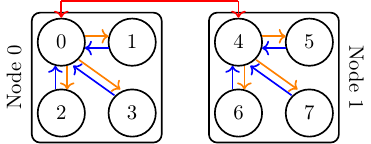}
\caption{Hierarchical all-to-all. Blue arrows indicate the intra-node gather, red arrows show the internode all-to-all operation, and yellow arrows indicate the intra-node scatter operation}
\label{fig:hierarchical}
\end{figure}

The hierarchical all-to-all, as described in Algorithm~\ref{alg:hierarchical} and corresponding Figure~\ref{fig:hierarchical}, 

\input{algs/hierarchical}
first gathers all data to a leader process, then performs all-to-all exchanges among leaders, and finally scatters data from the leader to all non-leader processes. These steps are shown in color in Algorithm~\ref{alg:hierarchical}, where these colors correspond to the arrows 
in~Figure~\ref{fig:hierarchical}.

This standard hierarchical exchange has exactly one leader per node, resulting in \texttt{local\_comm} containing all processes in a node and \texttt{group\_comm} containing the single leader per node.  This algorithm reduces the number of inter-node messages by limiting the number of leaders, minimizing contention of the network, as well as per-node resources such as the NIC.  However, with large numbers of active processes per node, the intra-node gather and scatter can result in significant overheads.

\begin{figure}[ht!]
\centering
\includegraphics[width=0.5\textwidth]{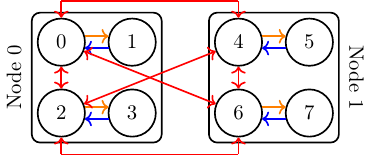}
\caption{Multi-leader all-to-all, 2 leaders per node. Blue arrows indicate the intra-node gather, red nodes show the all-to-all operation between leaders, and yellow arrows show the intra-node scatter operation.}
\label{fig:multileader}
\end{figure}

Multi-leader extensions to this algorithm consist of further partitioning the processes within each node so that multiple leaders are active within each all-to-all, as shown in Figure~\ref{fig:multileader}.
In the case of a multi-leader all-to-all, Algorithm~\ref{alg:hierarchical} receives a \texttt{local\_comm} communicator with the subset of processes per node corresponding to a single leader while \texttt{group\_comm} contains all leaders.  Multiple active leaders reduces the overheads of the gather and scatter operations.  However inter-leader all-to-all operation requires an exchange among multiple leaders per node, increasing the amount of inter-node communication.

\subsection{Node-Aware}
While hierarchical and multi-leader collectives reduce the number of inter-node messages, each message increases in size, and many processes sit idle during the all-to-all exchange between leaders.  

\input{algs/locality-aware}

\begin{figure}[ht!]
\centering
\includegraphics[width=0.5\textwidth]{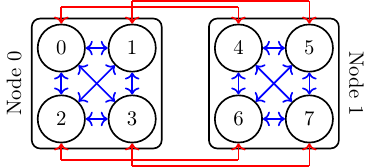}
\caption{Node-aware all-to-all. Red Arrows indicate the internode all-to-all operation, while blue arrows indicate the intra-node all-to-all operation.}
\label{fig:node-aware}
\end{figure}

Node-aware all-to-all exchanges, as detailed in Algorithm~\ref{alg:locality-aware} and corresponding Figure~\ref{fig:node-aware}, aggregate data among all processes per node, reducing the number of inter-node messages per process while evenly distributing data across all processes.  This exchange consists first performing an \texttt{MPI\_Alltoall} exchange on the \texttt{group\_comm} communicator (red arrows), which consists of processes across all nodes that have equal local rank.  After exchanging between nodes, all processes per node redistribute received data on \texttt{local\_comm} (blue arrows), the set of all processes per node. The node-aware algorithm greatly reduces inter-node communication costs for large exchanges in exchange for an intra-node redistribution of data.

\begin{figure}[ht!]
\centering
\includegraphics[width=0.5\textwidth]{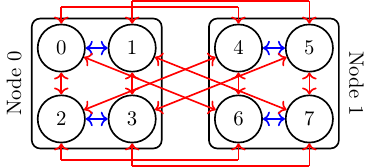}
\caption{Locality-aware all-to-all, 2 groups per node}
\label{fig:locality-aware}
\small Red arrows indicate the first round all-to-all, between regions, while blue arrows indicate the second all-to-all, within regions.
\end{figure}

On emerging many-core systems, the intra-node redistribution across all processes per node can create moderate overheads.  \textbf{This paper presents locality-aware aggregation, extending the standard node-aware approach with multiple groups of aggregation per node, as displayed in Figure~\ref{fig:locality-aware}.} Locality-aware aggregation is performed when \texttt{local\_comm} within Algorithm~\ref{alg:locality-aware} contains a subset of all processes per node, and \texttt{group\_comm} contains a single process from each subset with corresponding local rank.  

This optimization reduces the cost of redistributing data locally in exchange for a slight increase in steps of inter-node communication.

\subsection{Combining Multi-leader with Node-Aware}
While large bandwidth-bound exchanges are optimized when all processes communicate an equal portion of data, such as in locality-aware aggregation, small exchanges, which are bottlenecked by latency, benefit from the reductions in message count that are seen with hierarchical optimizations. 

\input{algs/multileader_locality}

However, multi-leader approaches increase inter-node message counts as each leader communicates with multiple leaders per node.  As core counts increase on nodes, the number of leaders in the multi-leader approach increases correspondingly to offset the overheads of intra-node gathers and scatters.  As a result, inter-node message counts are far from optimal.  

Note: MPICH also includes an alltoall implementation described as node-aware multi-leaders based, which is documented as: `Each rank on a node places the data for ranks sitting on other nodes into a shared memory buffer. Next each rank participates as a leader in inter-node Alltoall`.  The authors are unaware of any publication that further details this algorithm, but based on this MPICH documentation, this implementation requires all ranks to participate in inter-node communication, similar to the standard node-aware approach presented in Algorithm \ref{alg:locality-aware}.

\textbf{This paper presents a novel combination of multi-leader and node-aware approaches, as detailed in Algorithm~\ref{alg:hierarchical_locality}, in which the inter-node all-to-all exchange from Algorithm~\ref{alg:hierarchical} is replaced with the node-aware operation from Algorithm~\ref{alg:locality-aware}.} 

\begin{figure}[ht!]
\centering
\includegraphics[width=0.5\textwidth]{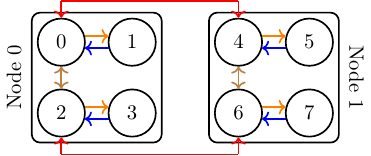}
\caption{Multi-leader + Node-Aware all-to-all, 2 leaders per node. Blue errors indicate intra-node gather operations to the leader, red arrows indicate the inter-node all-to-all operation, brown arrows indicate the inter-region all-to-all region, and yellow arrows show the intra-node scatter operations.}
\label{fig:multileader_loc}
\end{figure}

This multi-leader node-aware all-to-all, displayed in Figure~\ref{fig:multileader_loc} consists of first gathering to each leader (blue arrows) before exchanging data among corresponding leaders per node (red arrows).  Data is then redistributed among the leaders within each node (brown arrows) before finally being scattered to the non-leader processes (yellow arrows).  As a result, intra-node gather and scatter overheads are reduced with one leader per subset of each node's processes, while inter-node message counts remain low with each leader communicating only a single inter-node message to every node.

\section{Results}~\label{sec:results}
To analyze the performance of all existing and new node-aware all-to-all optimizations, we tested the algorithms across three many-core systems: Dane (at LLNL), Amber (at SNL), and Toulomne (LLNL). Table \ref{tab:architectures} describes the architectures of these systems. Dane and Amber consist of Intel Sapphire Rapids processors, with $112$ cores per node. System MPI on these machines is Intel MPI.  Toulomne uses AMD's MI-300A chips, which have $96$ cores per node. Toulomne uses Cray-MPICH as the system MPI implementation. On each system, we used all cores available per node.  All figures display the minimum of 3 runs for each data point.

\begin{table*}[ht]
\centering
\begin{tabular}{|l|l|l|l|l|}
\hline
\textbf{Name} & \textbf{CPU} & \textbf{Network} & \textbf{MPI} & \textbf{LibFabric Version}\\
\hline
Dane     & Intel Sapphire Rapids & Cornelis Networks Omni-Path & OpenMPI 4.1.2 & 2.2.0 \\
\hline
Amber    & Intel Sapphire Rapids & Cornelis Networks Omni-Path & OpenMPI 4.1.6 & 2.1.0 \\
\hline
Tuolomne & AMD Instinct MI300A   & Slingshot-11                & HPE Cray MPICH 8.1.32 & 2.1 \\
\hline
\end{tabular}
\caption{System Architectures}
\label{tab:architectures}
\end{table*}

Two experiments were performed on each system. The first experiment tested all-to-all algorithms using the maximum number of cores available per node ($112$ for Dane and Amber, $96$ on Tuolomne. All-to-all performance was then analyzed on all systems for data size exchanges from 4 to $4096$ bytes per process, and scaled from $2$ to $32$ nodes. While the per-message scales were limited to $4096$ bytes, at $32$ nodes ($3584$ processes on Dane and Amber), each process must exchange a buffer of $14,680,064$ bytes.  Additionally, the algorithms supporting multiple leaders (Multileader, Locality-Aware, and Multileader with Locality Awareness) were also tested with $4$, $8$, and $16$ processes per leader. It is important to note, none of the multileader or locality-aware aggregation groups were explicitly mapped to regions of locality, such as NUMA domains. In most cases, group sizes force the groups to cross NUMA regions and/or sockets. Mapping processes to specific cores is beyond the scope of this paper. However, while it would decrease the portability of the algorithms, explicitly forcing this mapping is likely to improve locality-aware and multi-leader approaches beyond the results presented in the remainder of this section.

The second experiment analyzed the performance of intra-node vs inter-node communication in the hierarchical, multi-leader, node-aware, locality-aware, and multileader with node-awareness approaches. In this experiment, each of the algorithms above were analyzed for performance for data size exchanges from $4$ to $4096$ bytes, scaled from $2$ to $32$ nodes. Additional timings were measured for any internal gathers, scatters, and all-to-alls within each algorithm. Additionally, multi-leader, locality-aware, and multi-leader with node awareness were evaluated with $4$, $8$, and $16$ processes per leader.  

All node-aware optimizations were compared against the system MPI all-to-all implementation.  
Further, each of Algorithms~\ref{alg:hierarchical} through~\ref{alg:hierarchical_locality} were tested with both 
pairwise exchange and non-blocking all-to-alls replacing each instance of \texttt{MPI\_Alltoall} within the algorithm.  
\textbf{Except in the case of System MPI, solid lines indicate all instances of \texttt{MPI\_Alltoall} calls using pairwise-exchange, while 
dashed lines indicate non-blocking underlying implementations.}  While System MPI results are also plotted as a solid line, note that 
since both IntelMPI and CrayMPICH are proprietary, the underlying algorithm in all cases is unknown. The System MPI performance is included in all plots as a point of reference.

\begin{figure}[ht!]
    \centering
    \includegraphics[width=0.8\linewidth]{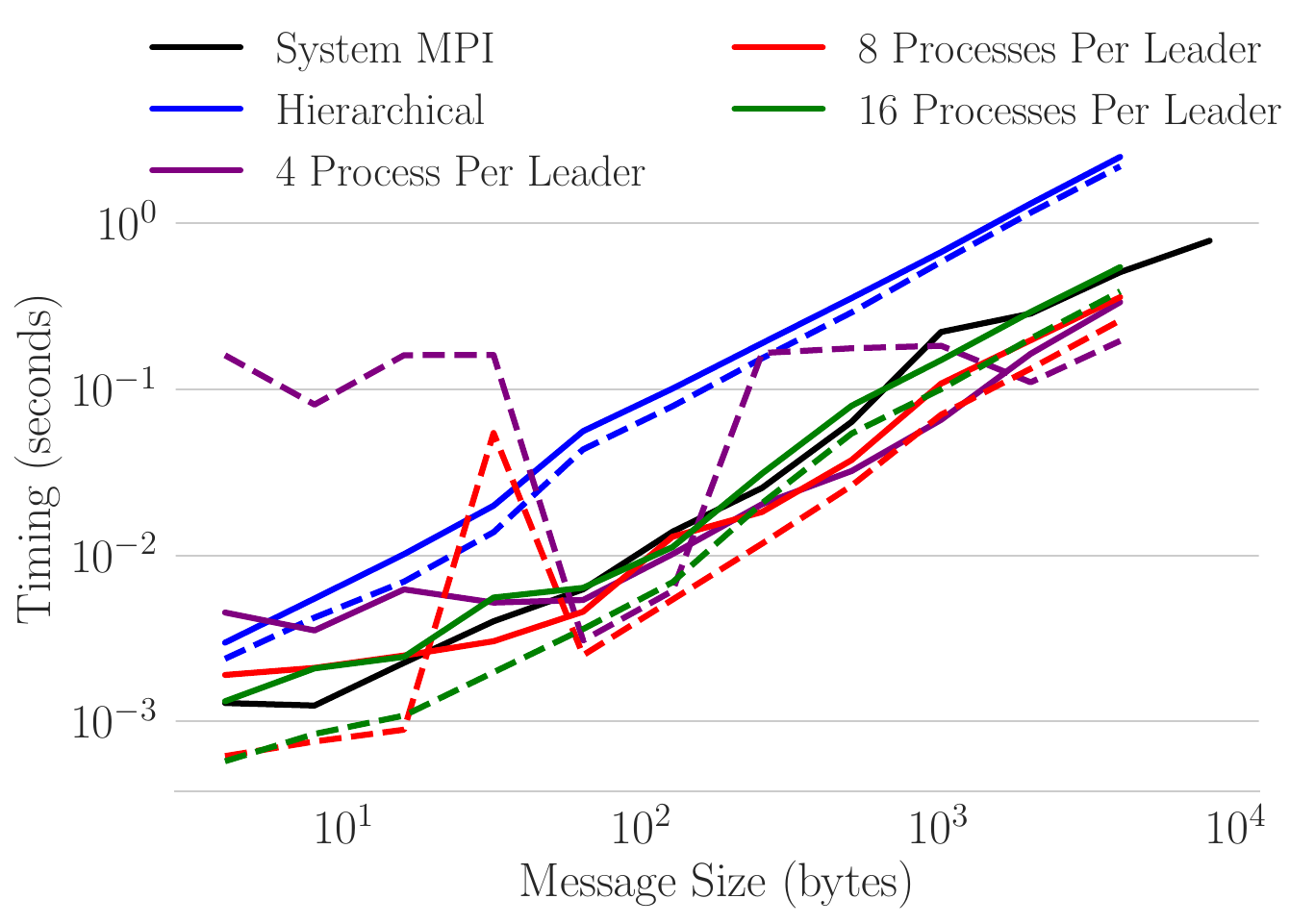}
    \caption{Hierarchical vs Multileader}
    \label{fig:hierarchical_multileader}
\end{figure}

\begin{figure}[ht!]
    \centering
    \includegraphics[width=0.8\linewidth]{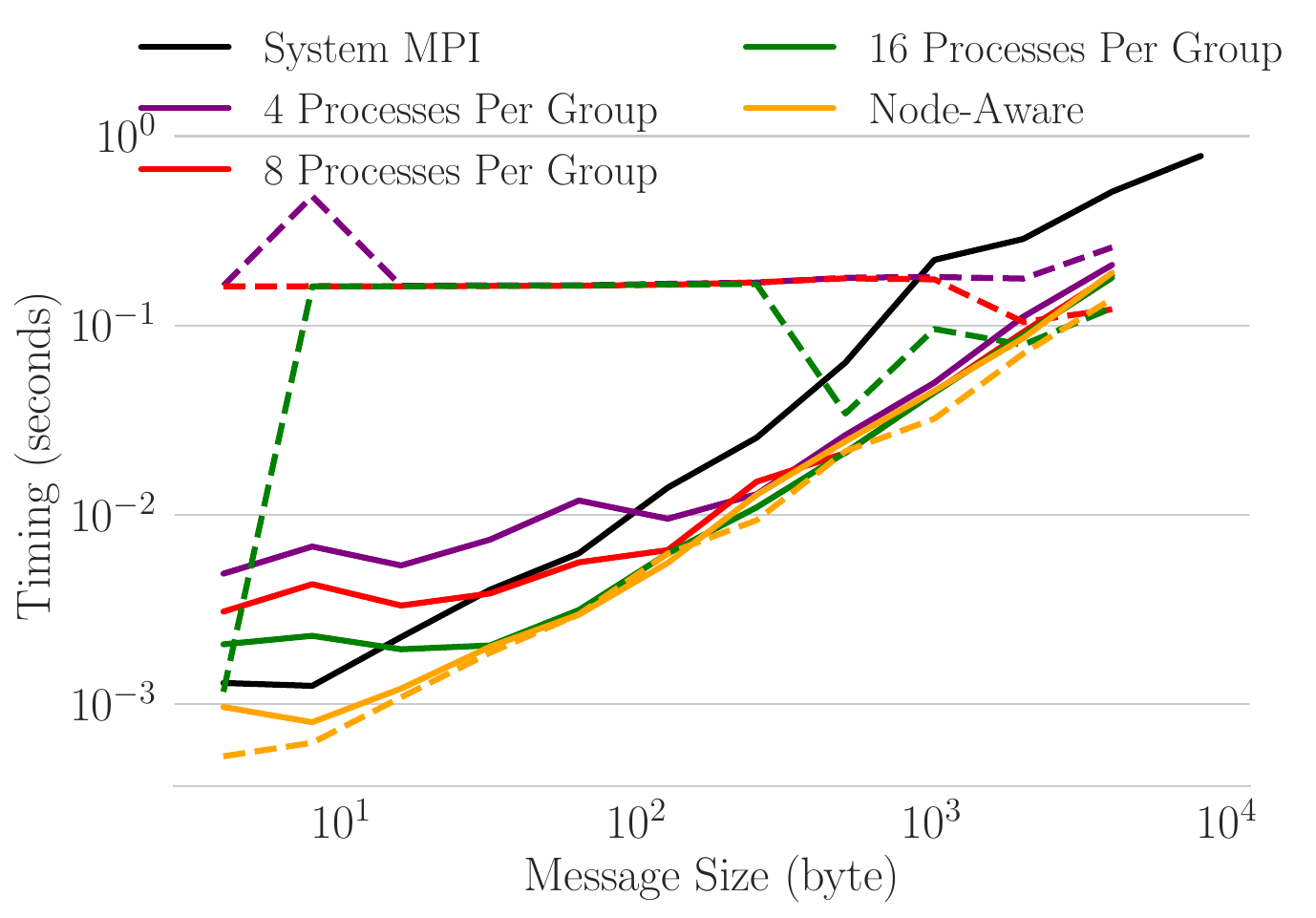}
    \caption{Node-Aware vs Locality-Aware}
    \label{fig:node_vs_locality}
\end{figure}

\begin{figure}[ht!]
    \centering
    \includegraphics[width=0.9\linewidth]{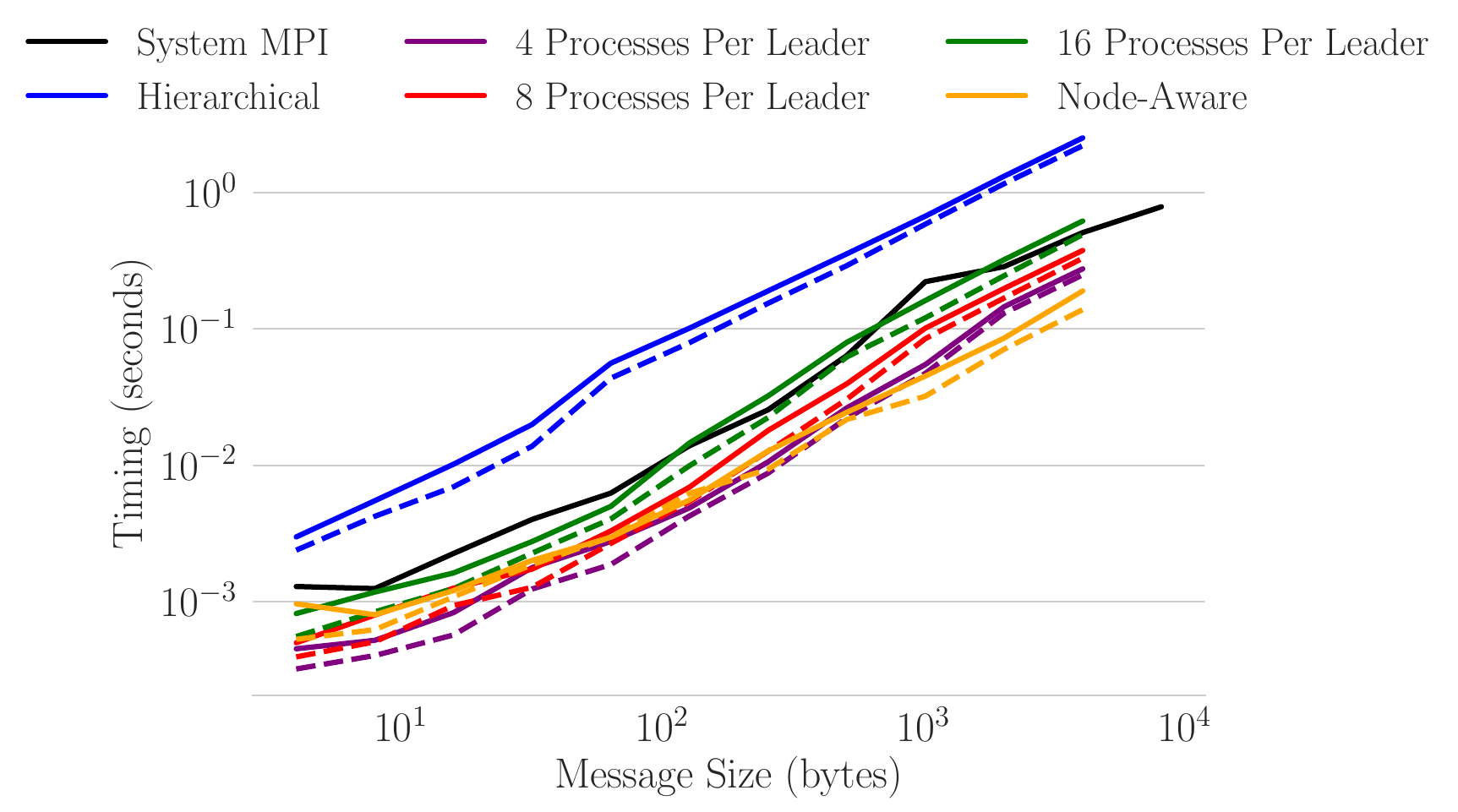}
    \caption{Multileader + Locality}
    \label{fig:multileader_locality}
\end{figure}

\begin{figure}[ht!]
    \centering
    \includegraphics[width=0.8\linewidth]{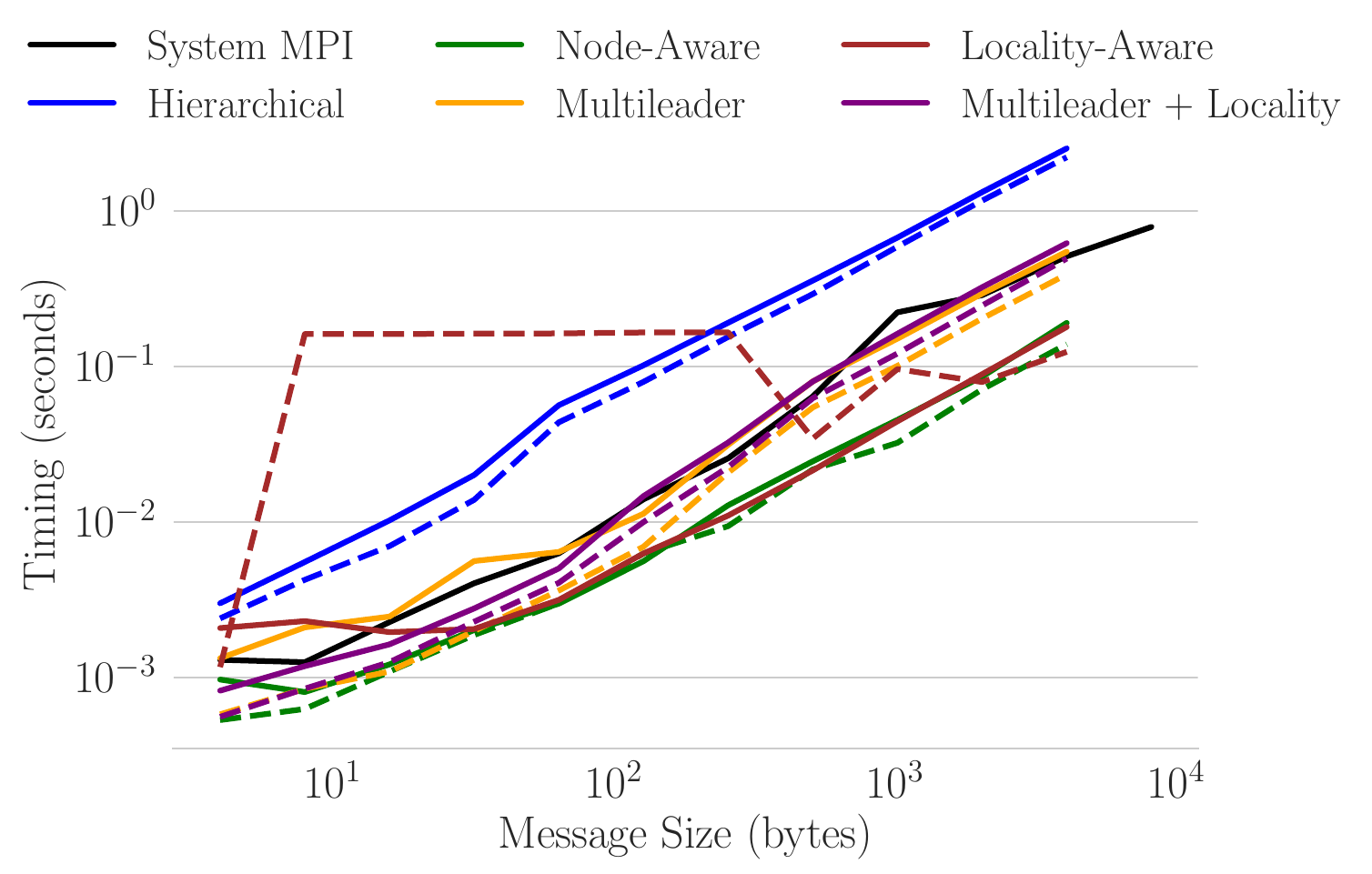}
    \caption{Various Sizes, 32 Nodes}
    \label{fig:dane_sizes32}
\end{figure}

\begin{figure}[ht!]
    \centering
    \includegraphics[width=0.8\linewidth]{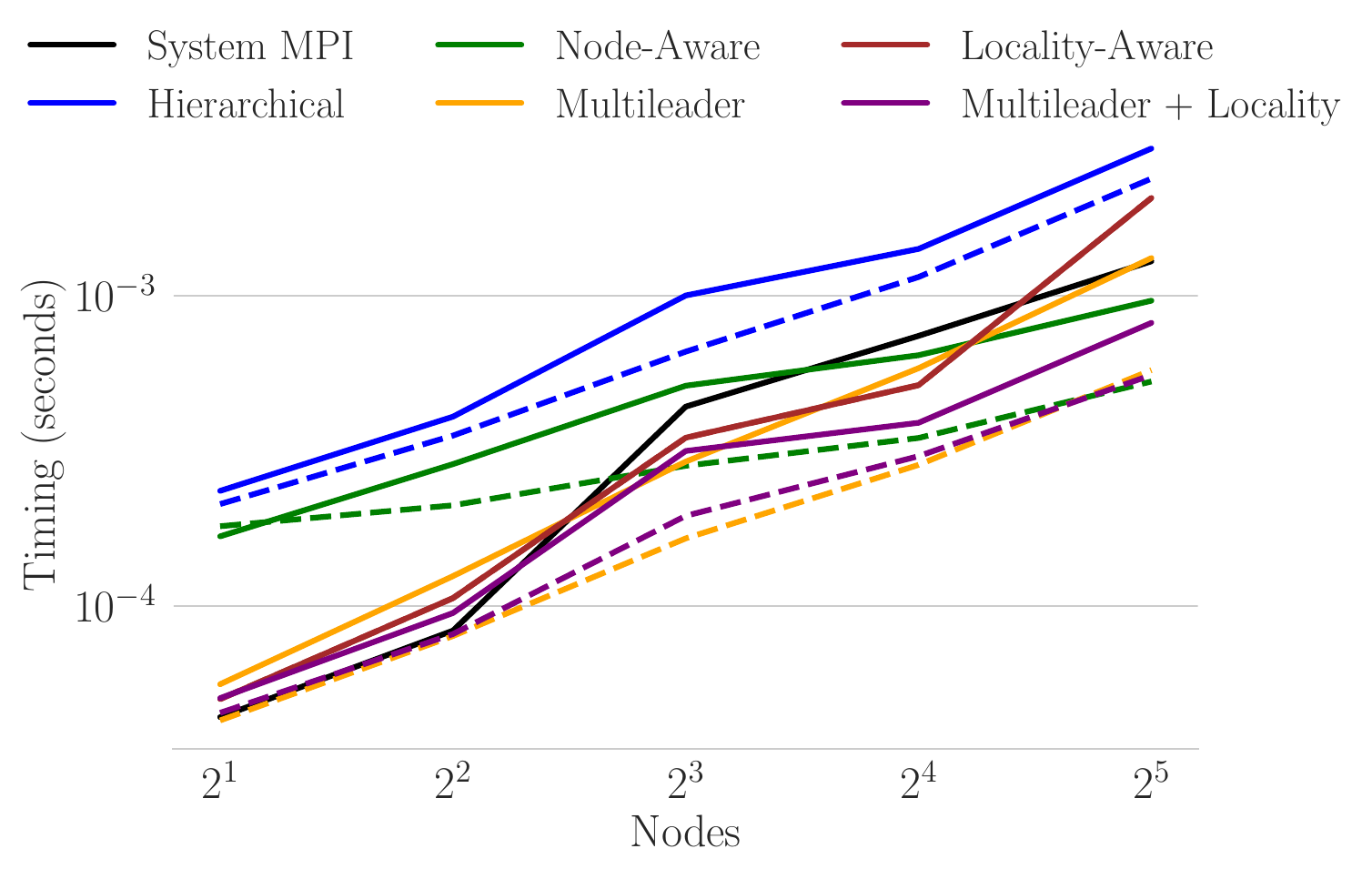}
    \caption{Message Size: 4 bytes, Node Scaling}
    \label{fig:dane_scale_small}
\end{figure}

\begin{figure}[ht!]
    \centering
    \includegraphics[width=0.8\linewidth]{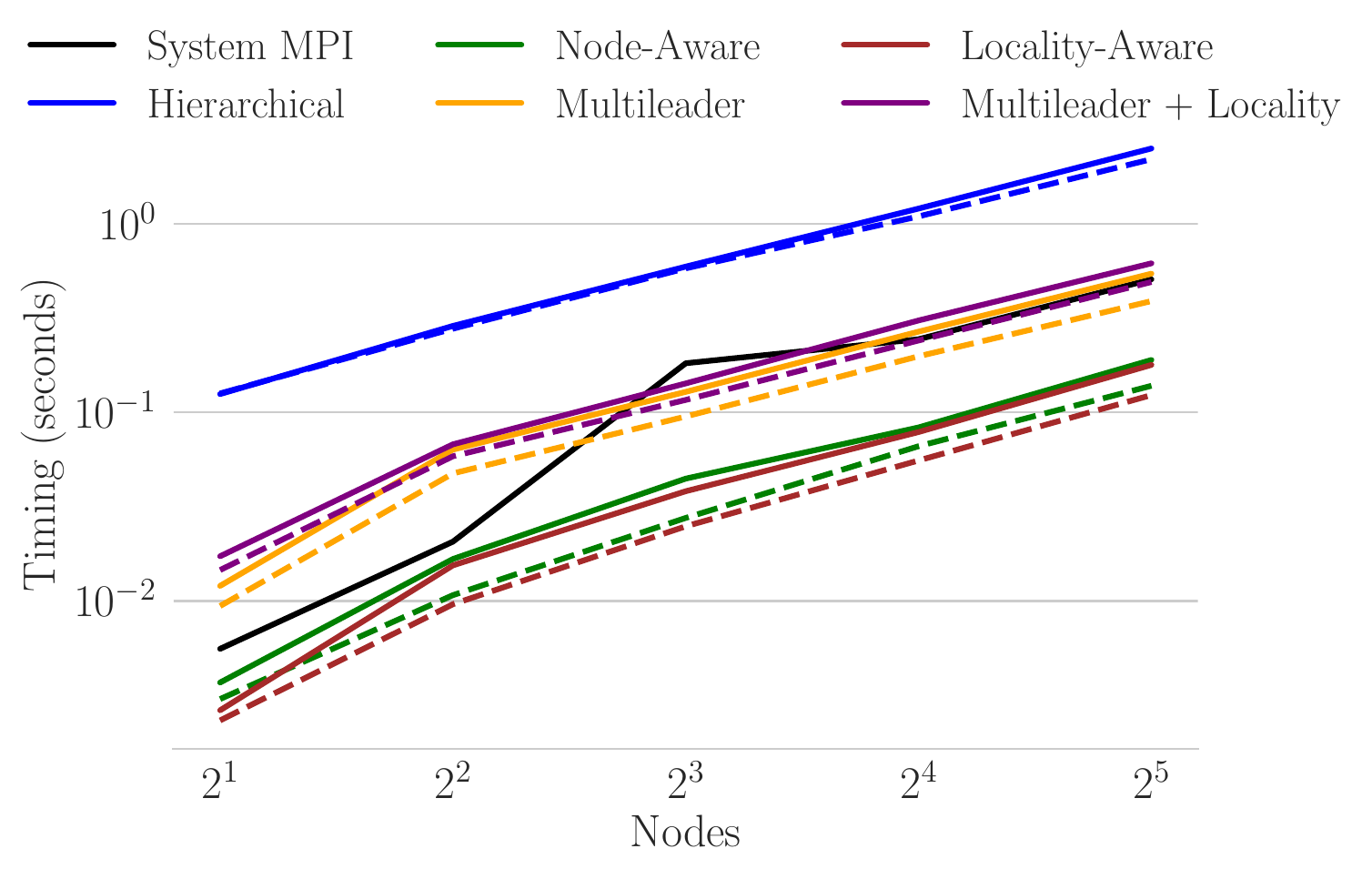}
    \caption{Message Size 4096 bytes, Node Scaling}
    \label{fig:dane_scale_large}
\end{figure}

Due to space constraints, figures~\ref{fig:hierarchical_multileader} through~\ref{fig:daneLocalityAwareProcScaling} analyze the performance of the various all-to-all algorithms only on Dane. The performance of the algorithms that are optimal at any message size is then presented for Amber and Tuolomne, in Figures~\ref{fig:amber_n32} and~\ref{fig:tuolomne_n32}, respectively.

Figure~\ref{fig:hierarchical_multileader} compares the performance of hierarchical all-to-all operations with multi-leader (Algorithm~\ref{alg:hierarchical}) for exchanges of a variety of sizes across 32 nodes of Dane. 
For large data sizes, the performance increases corresponding to the number of leaders per node, with the standard hierarchical all-to-all (e.g. a single leader per node) performing poorly.  Improvements are seen when increasing the number of leaders per node, decreasing the costs of the intra-node gather and scatter.  For smaller data sizes, multi-leader algorithms outperform standard hierarchical, but fewer leaders are beneficial, with $4$ leaders per node achieving best performance.  Intra-node gathers and scatters have less overhead for these smaller data sizes. Further, large numbers of leaders increase inter-node message counts.

Figure~\ref{fig:node_vs_locality} presents the cost of node- and locality-aware all-to-all exchanges (Algorithm~\ref{alg:locality-aware}) of various sizes across 32 nodes of Dane. While node-aware aggregation performs best for most data sizes, locality-aware aggregation with $4$ to $20$ groups outperforms node-aware aggregation at the largest tested data size.  While this improvement is only seen at the largest tested data size, the improvement is consistently seen from $8$ to $32$ nodes as shown in Figure~\ref{fig:dane_scale_large}. Note: we plot the minimum of each benchmarking run. The nonblocking all-to-all shows a large amount of variability, resulting in the sharp decrease in execution time at 1024 bytes in figures~\ref{fig:node_vs_locality} and ~\ref{fig:dane_sizes32}. Other runs do not show this dip.

Figure~\ref{fig:multileader_locality} shows the impact of the number of leaders per node on multi-leader node-aware all-to-alls (Algorithm~\ref{alg:hierarchical_locality}).  If a single leader was chosen, this algorithm reduces to the hierarchical approach.  If all processes per node are their own leaders, this reduces to the node-aware algorithm.  For small data sizes, this algorithm performs best with many but not all processes per node performing inter-node communication, limiting the number of leaders to around $20$.

Figure~\ref{fig:dane_sizes32} shows the performance of all algorithms for a variety of data sizes across $32$ nodes of Dane.  Based on the results presented in Figures~\ref{fig:hierarchical_multileader} to~\ref{fig:multileader_locality}, all multi-leader and locality-aware results use $28$ leaders/groups per node (4 processes per leader/group).  The multi-leader node-aware approach performs best for small data sizes, notable outperforming system MPI which is likely using the Bruck algorithm.  The node-aware algorithm optimizes larger data sizes while the largest sizes are improved with locality-aware aggregation.

Figures~\ref{fig:dane_scale_small} and~\ref{fig:dane_scale_large} show the performance of all algorithms when scaled across various node counts, for a data size of 4 bytes and 4096 bytes, respectively.

\subsection{Inter- and Intra-node communication}
\begin{figure}[ht!]
    \centering
    \includegraphics[width=0.8\linewidth]{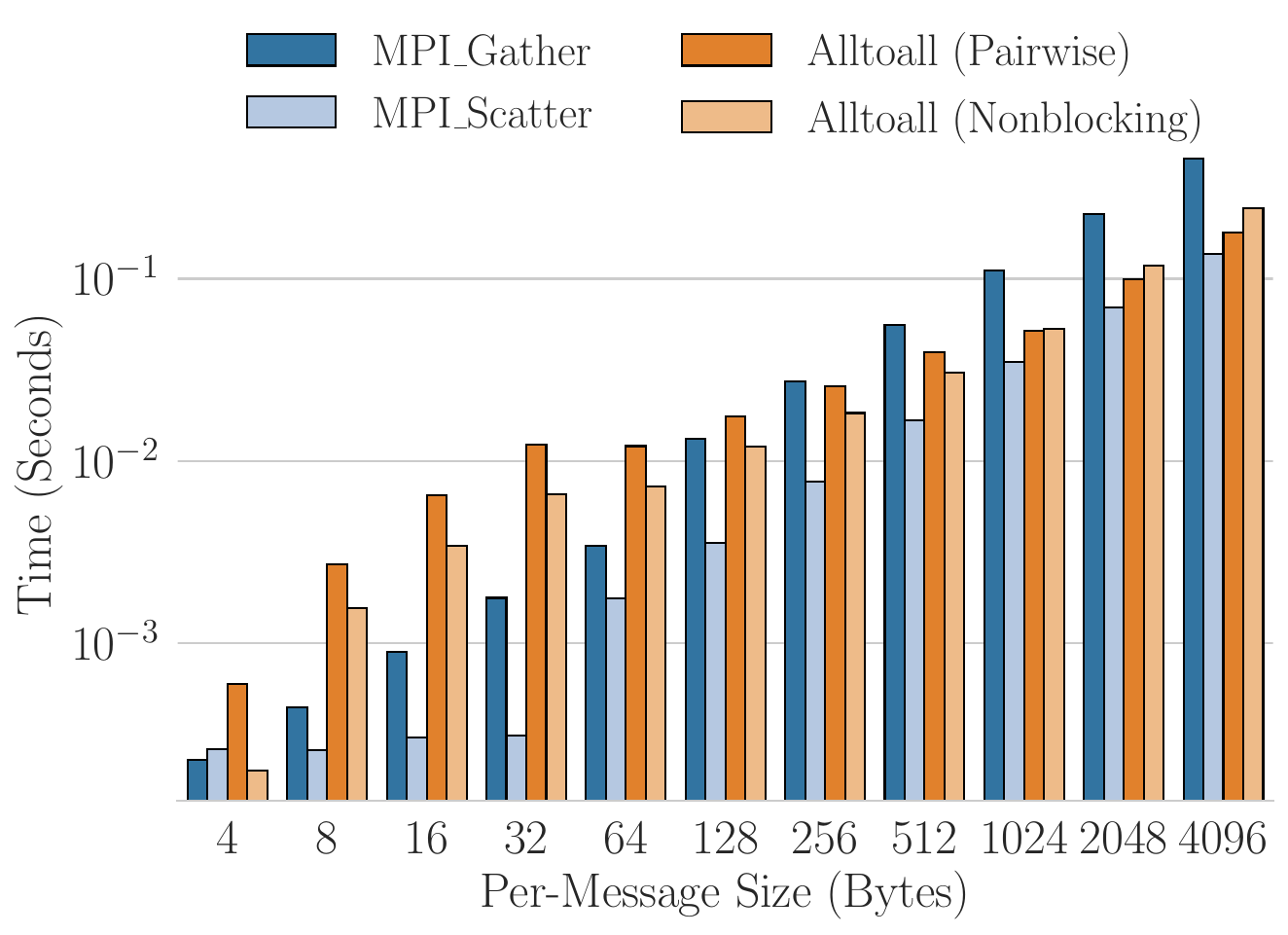}
    \caption{Dane, various sizes, 32 nodes, Hierarchical Timing Breakdown}
    \label{fig:danePairwiseHierarchical32}
\end{figure}

\begin{figure}[ht!]
    \centering
    \includegraphics[width=0.8\linewidth]{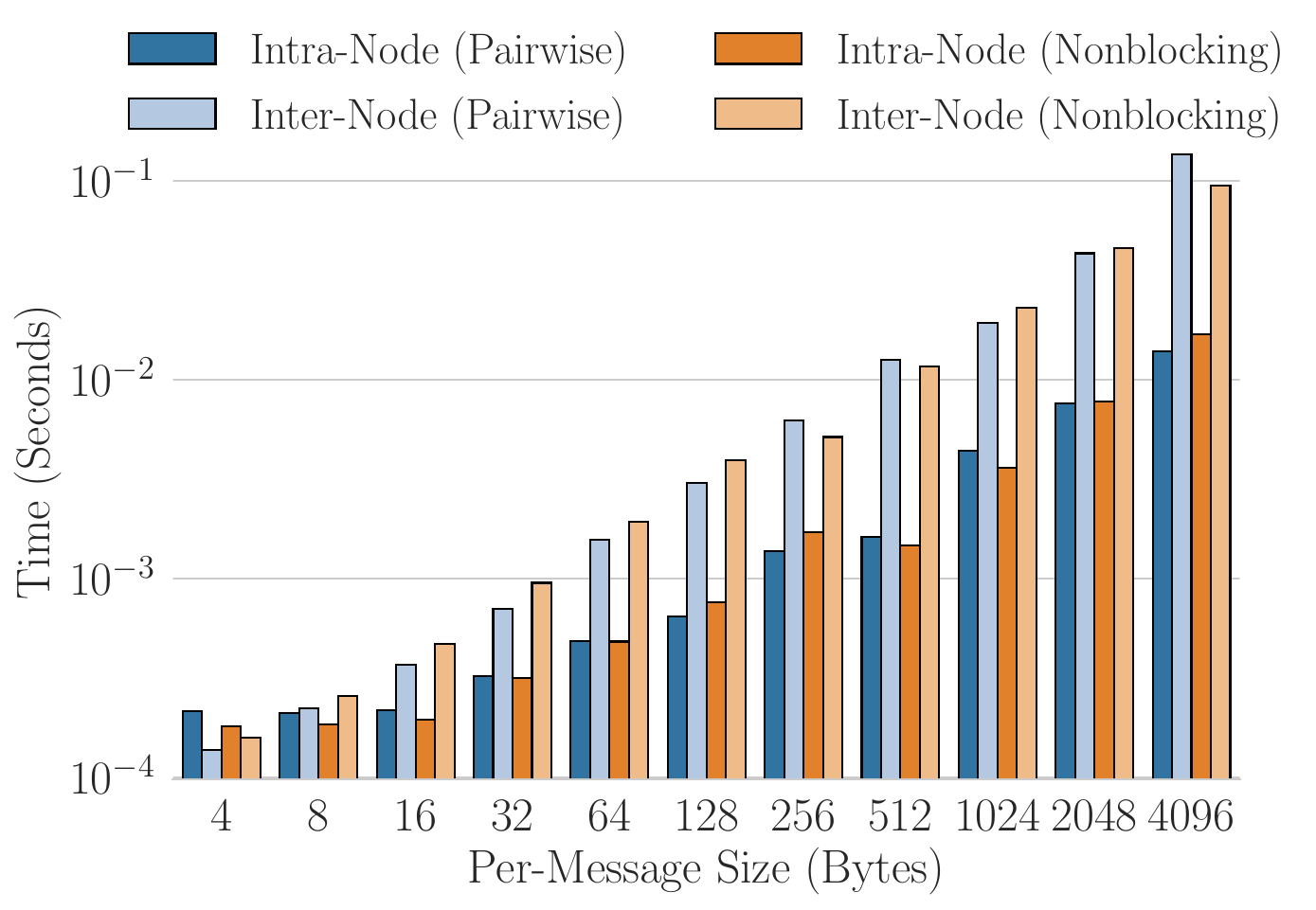}
    \caption{Dane, various sizes, 32 nodes, Node Aware Timing Breakdown}
    \label{fig:danePairwiseNodeAware32}
\end{figure}

\begin{figure}[ht!]
    \centering
    \includegraphics[width=0.8\linewidth]{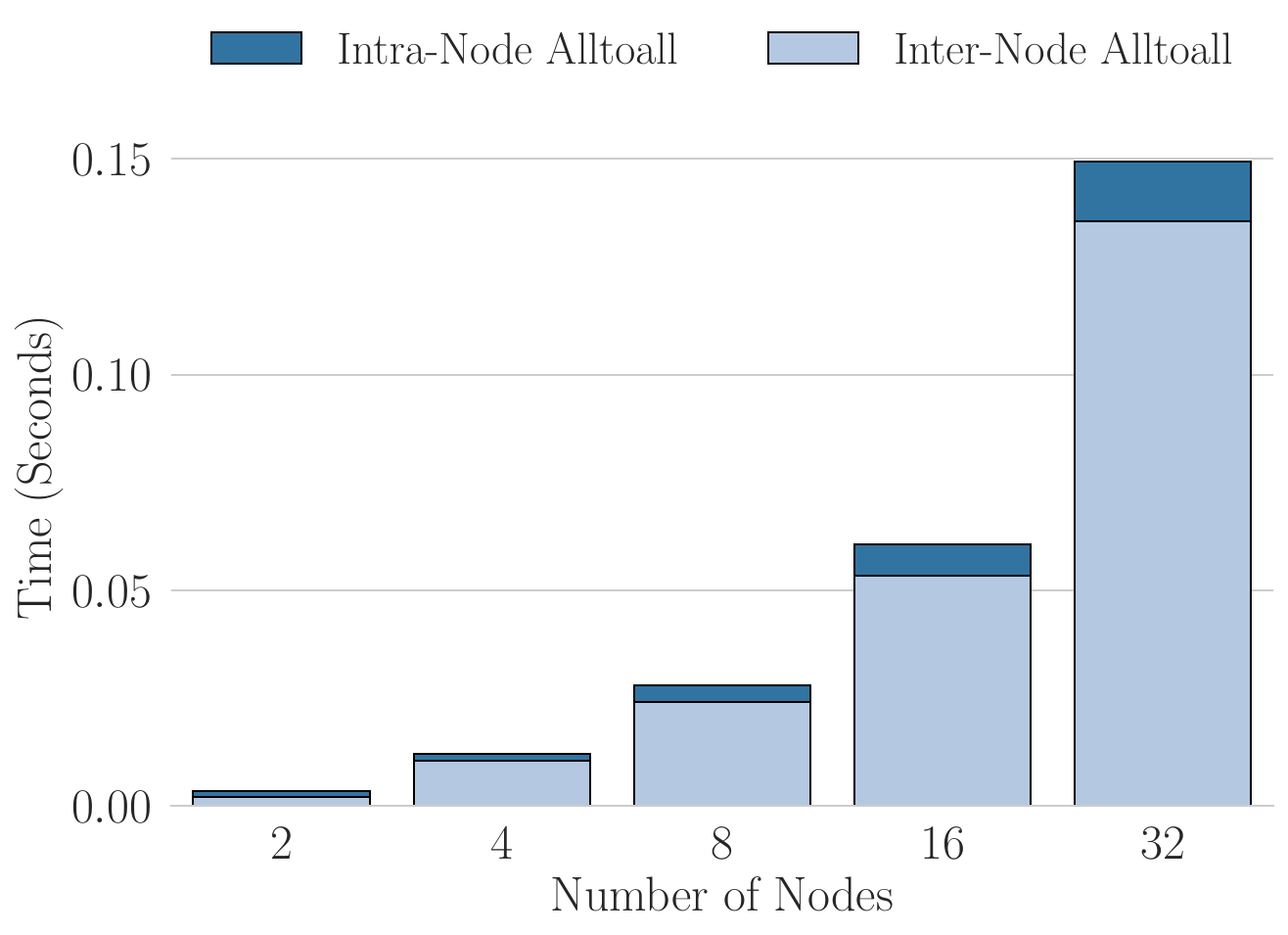}
    \caption{Dane, 1024 Integer Message Size, 2 to 32 nodes, Pairwise Node Aware}
    \label{fig:daneNodeAwareNodes}
\end{figure}

\begin{figure}[ht!]
    \centering
    \includegraphics[width=0.8\linewidth]{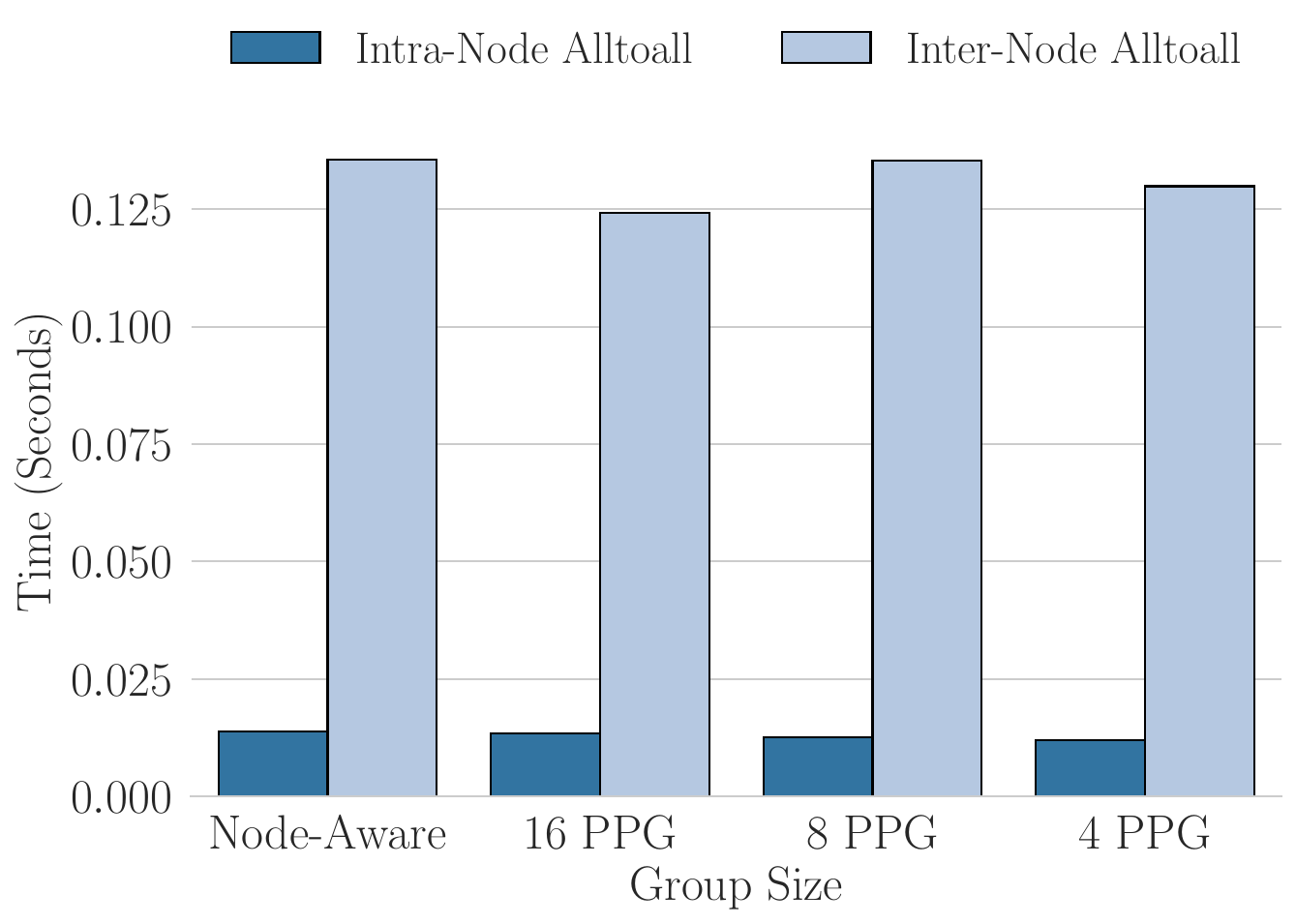}
    \caption{Dane, 1024 Integers, 32 Nodes, Various Processes per Leader, Locality Aware Timing Breakdown}
    \label{fig:daneLocalityAwareProcScaling}
\end{figure}

Due to space constraints, figures~\ref{fig:danePairwiseHierarchical32} through ~\ref{fig:daneLocalityAwareProcScaling} only show all-to-all algorithms that leverage pairwise exchange for the internal all-to-all operation. 
Figure~\ref{fig:danePairwiseHierarchical32} shows the performance of the individual communication components of the hierarchical algorithms on $32$ nodes on Dane. At smaller sizes (fewer than 256 byte messages), internode communication (the all-to-all operation between the single leader on each node) dominates communication. The nonblocking implementation outperforms the pairwise implementation until message sizes reach 2048 bytes. At larger message sizes (256 bytes and higher), the gather operation (intra-node communication) begins to dominate the execution time.

Figure ~\ref{fig:danePairwiseNodeAware32} shows the performance of the inter-node and intra-node communication in the node-aware algorithms on $32$ nodes on Dane. In both the pairwise and nonblocking cases, the inter-node communication again dominates the overall execution time. However, intra-node communication scales with internode communication.

Figure~\ref{fig:daneNodeAwareNodes} shows the performance of the individual components of the node-aware algorithm on Dane for a constant message size (4096 bytes), performed on node counts ranging from 2 to 32. Similar to the previous timing data, the inter-node communication dominates the overall latency, regardless of node count.

Figure~\ref{fig:daneLocalityAwareProcScaling} shows the impact of scaling the number of processes per group in the Locality-Aware algorithm, scaling from 1 process per group (the Node-Aware algorithm), to 16 processes per group.  In each case, the inter-node communication dominates the exchange, while the intra-node communication shows little variability between processes per group.  However, we do see slightly improved internode communication when there are 16 and 4 processes per group, which corresponds to 7 and 28 leaders, respectively, participating in the internode communication.

\subsection{Amber and Tuolomne}
\begin{figure}[ht!]
    \centering
    \includegraphics[width=0.8\linewidth]{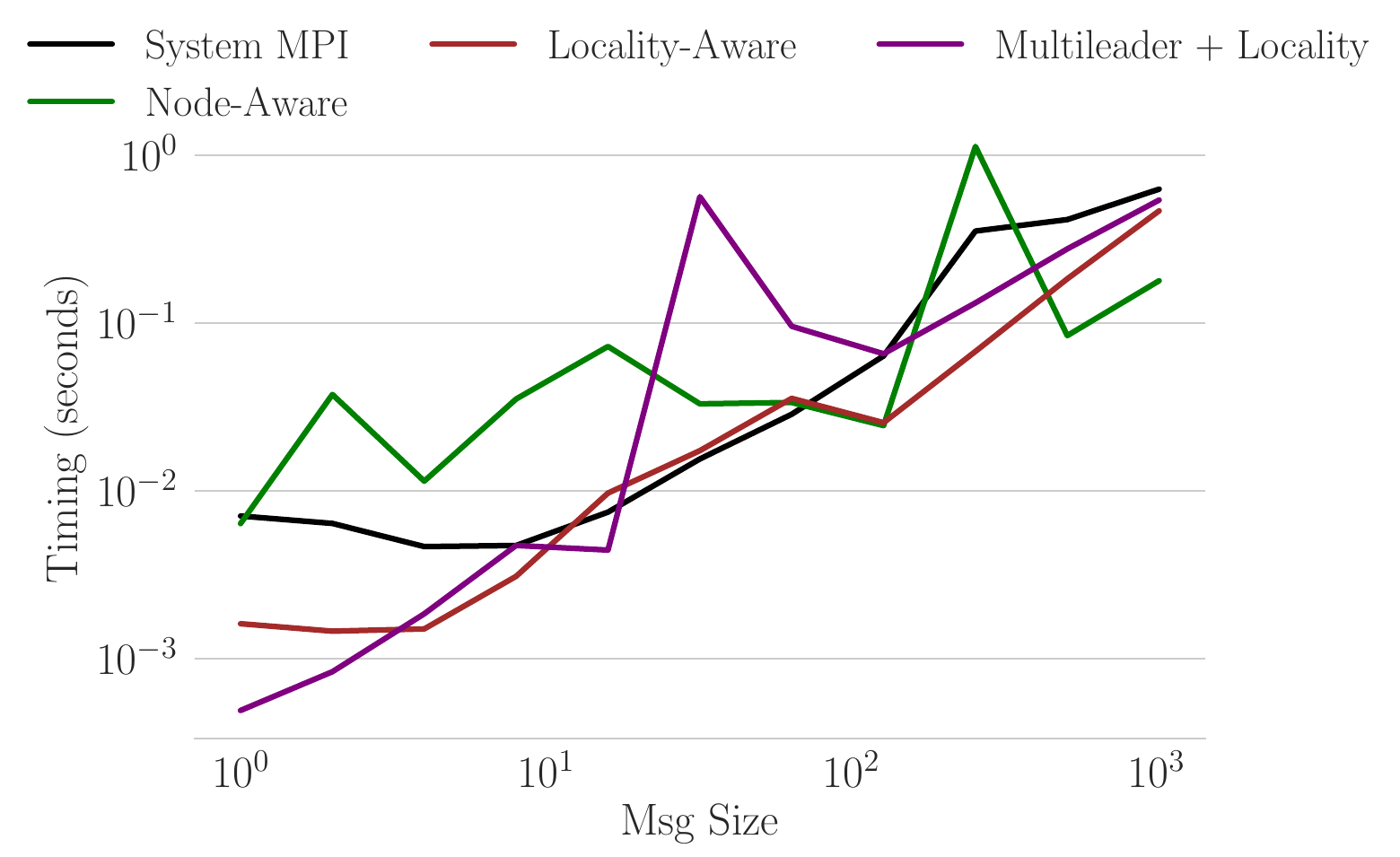}
    \caption{Amber, Various Sizes, 32 Nodes}
    \label{fig:amber_n32}
\end{figure}

Figure~\ref{fig:amber_n32} shows the performance of the node-aware, locality-aware, and multi-leader node-aware approaches, in comparison to system MPI, on 32 nodes of Amber.  Similar to previous results, the multi-leader node-aware approach performs best for small data sizes, while node-aware aggregation outperforms all other approaches for large data sizes.  

\begin{figure}[ht!]
    \centering
    \includegraphics[width=0.8\linewidth]{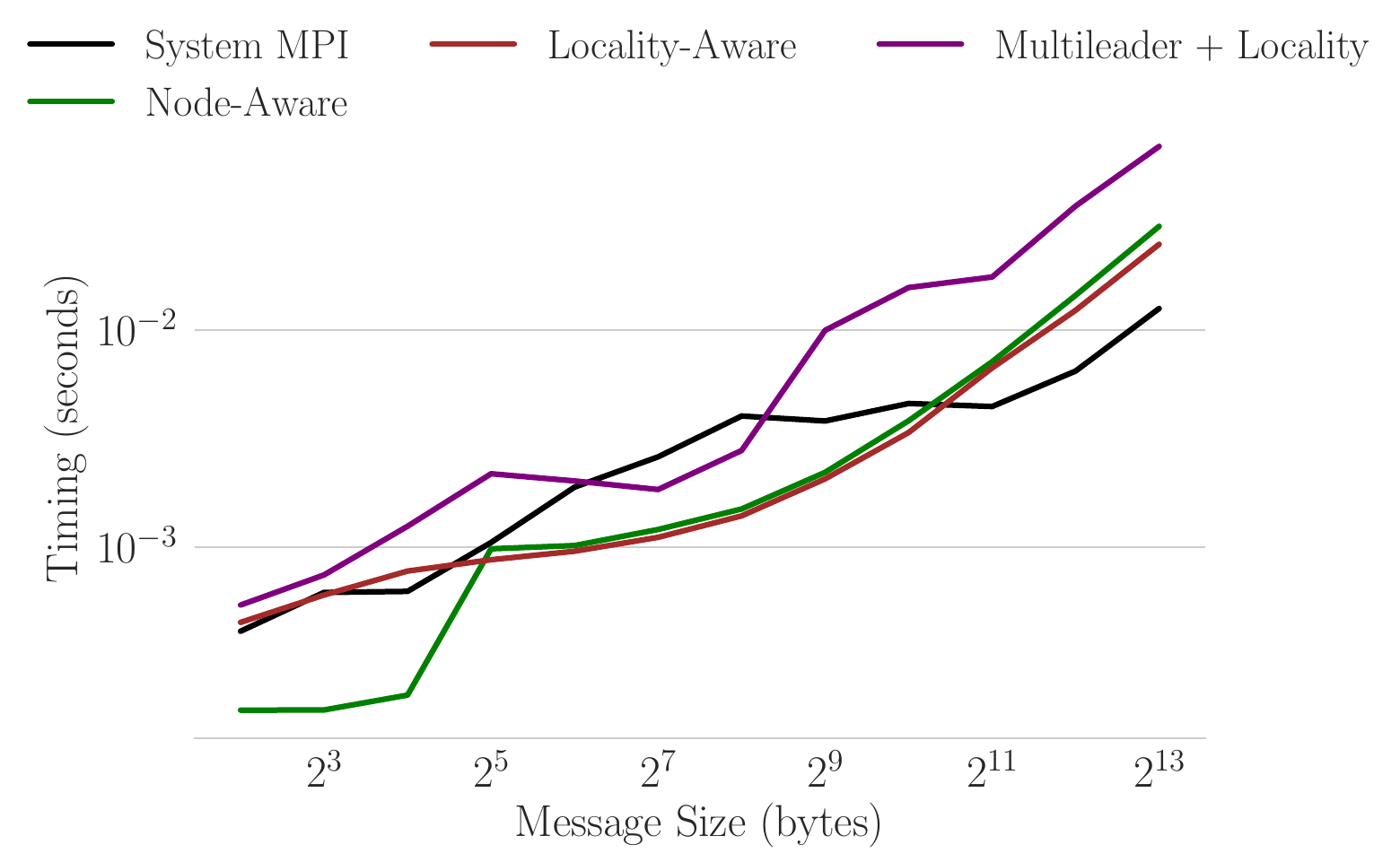}
    \caption{Tuolomne, Various Sizes, 32 Nodes}
    \label{fig:tuolomne_n32}
\end{figure}

Finally, figure~\ref{fig:tuolomne_n32} shows the performance of the all-to-all algorithms on 32 nodes of Tuolomne. In this case, the node-aware algorithm shows the best performance for small message sizes, though system MPI performance is only slightly slower. The locality-aware and multi-leader node-aware show greater latency. However, at larger message sizes, system MPI shows the best performance.

\subsection{Discussion}

The results exploring group and locality size in the Locality-Aware, Multileader, and Multileader with Node-Aware algorithms show an interesting trend. Figure~\ref{fig:daneLocalityAwareProcScaling} indicates that both 16 processes per group and 4 processes per group outperform other group sizes and the Node-Aware algorithm (in which all processes on a node are part of a single group), which indicates that optimizing the group/locality size is not a single-modal function. With the increasing prevalence of many-core nodes, multi-socket nodes with multiple NUMA domains per CPU, we posit that this is likely due to the added complexity of intra-node communication. 

\section{Future Work}~\label{sec:future}

The algorithms presented in this paper present opportunities for further exploration. One such area is extending the algorithms to other collective operations. This paper has focused on all-to-all which is a critical collective. We plan to extend this work by applying this approach on both other HPC critical collectives (all-gather, broadcast, etc.) and AI critical collectives (allreduce, reduce-scatter, etc.). Additionally, using the insights from the data gathered here, we can develop a model to evaluate these impacts at capability-scale. 

Future research includes the exploration the impact of NUMA domain on locality-aware algorithms, as well as the impact of using MPI Datatypes on the repacking cost inherent in these algorithms. Additionally, expanding this research to other systems, including those with GPUs, will be valuable in understanding how these algorithm choices can be applied more generally. Finally, we plan to explore how the optimal algorithm can be dynamically selected for a given computer, system MPI, process count, and data size. 

\section{Conclusions}~\label{sec:conclusions}
In this paper, we compare several all-to-all algorithms, including two novel algorithms, on Sapphire Rapids and MI-300A. We noted that the multi-leader node-aware approach showed superior performance to the other approaches for small message sizes across a range of node counts, outperforming system MPI in many cases.  Locality-aware aggregation shows promise, outperforming standard node-aware aggregation for the largest tested data sizes on Dane.  However, to achieve the full benefit of this method, groups should likely be mapped to regions of locality such as NUMA domains.  This paper also presented a study of the impact of group size and the number of leaders per node on multi-leader algorithms, showing generally improved performance with larger leader counts. Additionally, we demonstrated that inter-node communication dominates communication in hierarchical approaches for small message sizes, though intra-node communication dominates the data exchange for larger message sizes. However, in Node- and Locality-Aware exchanges, the inter-node communication dominates at all message sizes, regardless of node count.

We note that while leveraging non-blocking communication in the internal all-to-all operation often improves the speed at data exchanges, especially at smaller message sizes, it also shows greater variability that can result in unpredictable performance degradation. 

\begin{acks}
This work was performed with partial support from the National Science Foundation under Grant No. CCF-2338077 and the U.S. Department of Energy's National Nuclear Security Administration (NNSA) under the Predictive Science Academic Alliance Program (PSAAP-III), Award DE-NA0003966.

Any opinions, findings, and conclusions or recommendations expressed in this material are those of the authors and do not necessarily reflect the views of the National Science Foundation and the U.S. Department of Energy's National Nuclear Security Administration.

Sandia National Laboratories is a multi-mission laboratory managed and operated by National Technology \&  Engineering Solutions of Sandia, LLC (NTESS), a wholly owned subsidiary of Honeywell International Inc., for the U.S. Department of Energy’s National Nuclear Security Administration (DOE/NNSA) under contract DE-NA0003525. This written work is authored by an employee of NTESS. The employee, not NTESS, owns the right, title and interest in and to the written work and is responsible for its contents. Any subjective views or opinions that might be expressed in the written work do not necessarily represent the views of the U.S. Government. The publisher acknowledges that the U.S. Government retains a non-exclusive, paid-up, irrevocable, world-wide license to publish or reproduce the published form of this written work or allow others to do so, for U.S. Government purposes. The DOE will provide public access to results of federally sponsored research in accordance with the DOE Public Access Plan.

\end{acks}

\bibliographystyle{ACM-Reference-Format}
\bibliography{refs.bib}
\end{document}

%% file: algs/pairwise.tex
\begin{algorithm2e}[ht!]
  \DontPrintSemicolon%
  \KwIn{$p$\tcc*{process rank}
        $n$\tcc*{process count}
        $s_{\texttt{size}}$, $s_{\texttt{type}}$, $s_{\texttt{buf}}$\tcc*{send size, type, and buffer}
        $r_{\texttt{size}}$, $r_{\texttt{type}}$, $r_{\texttt{buf}}$\tcc*{recv size, type, and buffer}
        }

  \BlankLine%
  
  \For{$i\gets0$ \KwTo $n$}{
      $s_{\texttt{proc}} = p + i \mod n$\;
      $r_{\texttt{proc}} = p + n - i \mod n$\;
      \texttt{MPI\_Sendrecv}($s_{\texttt{buf}}$,$s_{\texttt{size}}$,$s_{\texttt{type}}$,$s_{\texttt{proc}}$,\ldots)\,$r_{\texttt{buf}}$,$r_{\texttt{size}}$,$r_{\texttt{type}}$,$r_{\texttt{proc}}$,\ldots)\;
  }

    \caption{Pairwise Exchange}\label{alg:pairwise_exchange}
\end{algorithm2e}

%% file: algs/nonblocking.tex
\begin{algorithm2e}[ht!]
  \DontPrintSemicolon%
  \KwIn{$p$\tcc*{process rank}
        $n$\tcc*{process count}
        $s_{\texttt{size}}$, $s_{\texttt{type}}$, $s_{\texttt{buf}}$\tcc*{send size, type, and buffer}
        $r_{\texttt{size}}$, $r_{\texttt{type}}$, $r_{\texttt{buf}}$\tcc*{recv size, type, and buffer}
        }

  \BlankLine%
  
  \For{$i\gets0$ \KwTo $n$}{
      $s_{\texttt{proc}} = p + i \mod n$\;
      $r_{\texttt{proc}} = p + n - i \mod n$\;
      \texttt{MPI\_Isend}($s_{\texttt{buf}}$,$s_{\texttt{size}}$,$s_{\texttt{type}}$,$s_{\texttt{proc}}$,\ldots)\;
      \texttt{MPI\_Irecv}($r_{\texttt{buf}}$,$r_{\texttt{size}}$,$r_{\texttt{type}}$,$r_{\texttt{proc}}$,\ldots)\;
  }

   \texttt{MPI\_Waitall}($2\times (n-1)$, \ldots)\;
  
    \caption{Non-blocking}\label{alg:nonblocking}
\end{algorithm2e}

%% file: algs/hierarchical.tex
\begin{algorithm2e}[ht!]
  \DontPrintSemicolon%
  \KwIn{$p$\tcc*{process rank}
        $n$\tcc*{process count}
        $s_{\texttt{size}}$, $s_{\texttt{type}}$, $s_{\texttt{buf}}$\tcc*{send size, type, and buffer}
        $r_{\texttt{size}}$, $r_{\texttt{type}}$, $r_{\texttt{buf}}$\tcc*{recv size, type, and buffer}
        \texttt{local\_comm}\tcc*{All processes local to region}
        $ppn$, $l$\tcc*{Size and rank of local\_comm}
        \texttt{group\_comm}\tcc*{All processes with equal local rank}
        }

  \BlankLine%

  $s_{\texttt{buf}_{leader}} \leftarrow \texttt{buffer of size }s_{\texttt{size}} \cdot n \cdot ppn$\;
  $r_{\texttt{buf}_{leader}} \leftarrow \texttt{buffer of size }r_{\texttt{size}} \cdot n \cdot ppn$\;
  \BlankLine%

  \tcp{Gather to leader}
  \textcolor{blue}{MPI\_Gather($s_{\texttt{buf}}$, $s_{\texttt{size}}\cdot n$, \ldots , $s_{\texttt{buf}_{leader}}$, \ldots, \texttt{local\_comm})}\;
  
  \BlankLine%
  \texttt{Repack Data}\;
  \BlankLine%

  \tcp{Alltoall exchange between leaders}
  \textcolor{red}{MPI\_Alltoall($s_{\texttt{buf}_{leader}}$, $s_{\texttt{size}}\cdot ppn^{2}$, \ldots,  $r_{\texttt{buf}_{leader}}$, $r_{\texttt{size}}\cdot ppn^{2}$\ldots, \texttt{group\_comm})}\;

  \BlankLine%
  \texttt{Repack Data}\;
  \BlankLine%

  \tcp{Scatter from leader}
  \textcolor{orange}{MPI\_Scatter($r_{\texttt{buf}_{leader}}$, $r_{\texttt{size}}\cdot n$, \ldots, $r_{\texttt{buf}}$, \ldots, \texttt{local\_comm})}\;

    \caption{Hierarchical}\label{alg:hierarchical}
\end{algorithm2e}

%% file: algs/locality-aware.tex
\begin{algorithm2e}[ht!]
  \DontPrintSemicolon%
  \KwIn{$p$\tcc*{process rank}
        $n$\tcc*{process count}
        $s_{\texttt{size}}$, $s_{\texttt{type}}$, $s_{\texttt{buf}}$\tcc*{send size, type, and buffer}
        $r_{\texttt{size}}$, $r_{\texttt{type}}$, $r_{\texttt{buf}}$\tcc*{recv size, type, and buffer}
        \texttt{local\_comm}\tcc*{All processes local to region}
        $ppn$, $l$\tcc*{Size and rank of local\_comm}
        \texttt{group\_comm}\tcc*{All processes with equal local rank}
        }

  \BlankLine%

  $tmp_{\texttt{buf}} \leftarrow \texttt{buffer of size }s_{\texttt{size}}$\;
  \BlankLine%

  \tcp{Inter-region Alltoall}
  \textcolor{red}{MPI\_Alltoall($s_{\texttt{buf}}$, $s_{\texttt{size}}\cdot ppn$, \ldots,  $tmp_{\texttt{buf}}$, $r_{\texttt{size}}\cdot ppn$\ldots, \texttt{group\_comm})}\;

  \BlankLine%
  \texttt{Repack Data}\;
  \BlankLine%

  \tcp{Intra-region Alltoall}
  \textcolor{blue}{MPI\_Alltoall($tmp_{\texttt{buf}}$, $r_{\texttt{size}}\cdot ppn$, \ldots,  $r_{\texttt{buf}}$, $r_{\texttt{size}}\cdot ppn$\ldots, \texttt{local\_comm})}\;

    \caption{Node-Aware}\label{alg:locality-aware}
\end{algorithm2e}

%% file: algs/multileader_locality.tex
\begin{algorithm2e}[ht!]
  \DontPrintSemicolon
  \KwIn{$p$\tcc*{process rank}
        $n$\tcc*{process count}
        $s_{\texttt{size}}$, $s_{\texttt{type}}$, $s_{\texttt{buf}}$\tcc*{send size, type, and buffer}
        $r_{\texttt{size}}$, $r_{\texttt{type}}$, $r_{\texttt{buf}}$\tcc*{recv size, type, and buffer}
        \texttt{node\_comm}\tcc*{All processes local to node}
        $ppn$, $l$\tcc*{Size and rank of node\_comm}
        \texttt{leader\_comm}\tcc*{All processes local to leader}
        $ppl$\tcc*{Size of leader\_comm}
        \texttt{group\_comm}\tcc*{All processes with equal local rank $l$}
        $n_{nodes}$\tcc*{size of group\_comm}
        \texttt{leader\_group\_comm}\tcc*{All leaders local to the same node}
        }

  \BlankLine%

  $s_{\texttt{buf}_{leader}} \leftarrow \texttt{buffer of size }s_{\texttt{size}} \cdot n \cdot ppn$\;
  $r_{\texttt{buf}_{leader}} \leftarrow \texttt{buffer of size }r_{\texttt{size}} \cdot n \cdot ppn$\;
  \BlankLine%

  \tcp{Gather to leader}
  \textcolor{blue}{MPI\_Gather($s_{\texttt{buf}}$, $s_{\texttt{size}}\cdot n$, \ldots , $s_{\texttt{buf}_{leader}}$, \ldots, \texttt{local\_comm})}\;
  
  \BlankLine%
  \texttt{Repack Data}\;
  \BlankLine%

  \tcp{Inter-region Alltoall}
  \textcolor{red}{MPI\_Alltoall($s_{\texttt{buf}_{leader}}$, $s_{\texttt{size}}\cdot ppn \cdot ppl$, \ldots,  $r_{\texttt{buf}_{leader}}$, $r_{\texttt{size}}\cdot ppn \cdot ppl$\ldots, \texttt{group\_comm})}\;

  \BlankLine%
  \texttt{Repack Data}\;
  \BlankLine%

  \tcp{Intra-region Alltoall}
  \textcolor{brown}{MPI\_Alltoall($r_{\texttt{buf}_{leader}}$, $r_{\texttt{size}}\cdot n_{nodes} \cdot ppl^{2}$, \ldots,  $s_{\texttt{buf}_{leader}}$, $r_{\texttt{size}}\cdot n_{nodes} \cdot ppl^{2}$, \ldots, \texttt{leader\_group\_comm})}\;

  \BlankLine%
  \texttt{Repack Data}\;
  \BlankLine%

  \tcp{Scatter from leader}
  \textcolor{orange}{MPI\_Scatter($s_{\texttt{buf}_{leader}}$, $r_{\texttt{size}}\cdot n$, \ldots, $r_{\texttt{buf}}$, \ldots, \texttt{local\_comm})}\;

    \caption{Multi-Leader + Node-Aware}\label{alg:hierarchical_locality}
\end{algorithm2e}